\documentclass[12pt,a4paper]{article}

\usepackage[T1]{fontenc}
\usepackage[utf8]{inputenc}
\usepackage{lmodern}
\usepackage{microtype}
\usepackage[a4paper,margin=2.5cm]{geometry}
\usepackage{setspace}
\usepackage{amsmath,amssymb}
\usepackage{siunitx}
\usepackage{graphicx}
\graphicspath{{figures/}}
\usepackage{booktabs}
\usepackage{array}
\usepackage{caption}
\usepackage{authblk}
\usepackage[numbers,sort&compress]{natbib}
\usepackage[hidelinks]{hyperref}
\usepackage[nameinlink,noabbrev]{cleveref}
\crefname{figure}{Fig.}{Figs.}
\Crefname{figure}{Figure}{Figures}
\crefname{table}{Table}{Tables}
\Crefname{table}{Table}{Tables}
\crefname{section}{Sect.}{Sects.}
\Crefname{section}{Section}{Sections}
\crefname{equation}{}{}

\title{Kaonic Atom Measurements with the SIDDHARTA-2 Experiment at the DA$\Phi$NE accelerator}
\author[1,2,3]{D~Sirghi$^{*}$}
\author[4,2]{L~Abbene}
\author[2,6]{F~Artibani}
\author[2]{M~Bazzi}
\author[7,8]{G~Borghi}
\author[9]{D~Bosnar}
\author[3]{M~Bragadireanu}
\author[4,2]{A~Buttacavoli}
\author[7,8]{M~Carminati}
\author[2]{A~Clozza}
\author[2,10]{F~Clozza}
\author[2]{L~De~Paolis}
\author[11,2]{R~Del~Grande}
\author[2,12,13]{K~Dulski}
\author[7,8]{C~Fiorini}
\author[9]{I~Friščić}
\author[2]{C~Guaraldo$^{\dagger}$}
\author[2]{M~Iliescu}
\author[14]{M~Iwasaki}
\author[12,13,2]{A~Khreptak}
\author[2]{S~Manti}
\author[5]{J~Marton$^{**}$}
\author[12,13]{P~Moskal}
\author[15]{H~Ohnishi}
\author[1,2]{K~Piscicchia}
\author[4,2]{F~Principato}
\author[2]{A~Scordo}
\author[2]{F~Sgaramella}
\author[12]{M~Silarski}
\author[2,3]{F~Sirghi}
\author[12,13,2]{M~Skurzok}
\author[2]{A~Spallone}
\author[15,2]{K~Toho}
\author[5]{J~Zmeskal$^{\dagger}$}
\author[2,3]{C~Curceanu}

\affil[1]{Centro Ricerche Enrico Fermi, Museo Storico della Fisica e Centro Studi e Ricerche ``Enrico Fermi'', Roma, Italy}
\affil[2]{Laboratori Nazionali di Frascati INFN, Frascati, Italy}
\affil[3]{IFIN-HH, Institutul National Pentru Fizica si Inginerie Nucleara Horia Hulubei, 30 Reactorului, 077125, Magurele, Romania}
\affil[4]{Department of Physics and Chemistry (DiFC), Emilio Segrè, University of Palermo, Palermo, Italy}
\affil[5]{Stefan Meyer Institute for Subatomic Physics, Vienna, Austria}
\affil[6]{Università degli Studi di Roma Tre, Dipartimento di Fisica, Roma, Italy}
\affil[7]{Politecnico di Milano, Dipartimento di Elettronica, Informazione e Bioingegneria, Milano, Italy}
\affil[8]{INFN Sezione di Milano, Milano, Italy}
\affil[9]{Department of Physics, Faculty of Science, University of Zagreb, Zagreb, Croatia}
\affil[10]{Università degli Studi di Roma Tor Vergata, Dipartimento di Fisica, Roma, Italy}
\affil[11]{Faculty of Nuclear Sciences and Physical Engineering, Czech Technical University in Prague, Břehovà 7, 115 19, Prague, Czech Republic}
\affil[12]{Faculty of Physics, Astronomy, and Applied Computer Science, Jagiellonian University, Kraków, Poland}
\affil[13]{Center for Theranostics, Jagiellonian University, Krakow, Poland}
\affil[14]{RIKEN, Tokyo, Japan}
\affil[15]{Research Center for Accelerator and Radioisotope Science (RARiS), Tohoku University, Sendai, Japan}

\date{}

\begin{document}
\maketitle

\begin{center}
\footnotesize
$^\dagger$ Deceased\\
$^*$ Diana.Laura.Sirghi@lnf.infn.it (Corresponding Author)\\
$^{**}$ Now at Atominstitut, Technische Universit\"at Wien, 1020, Vienna, Austria
\end{center}

\vspace{6pt}

\begin{abstract}
Kaonic atoms provide a unique experimental tool to investigate low-energy strong interactions in systems containing strangeness and, through high-lying transitions, to test bound-state quantum electrodynamics. The SIDDHARTA-2 experiment at the DA$\Phi$NE electron--positron collider performs high-precision X-ray spectroscopy of kaonic atoms using low-momentum charged kaons produced in $\phi$-meson decays. Its primary goal is the study of kaonic deuterium through the measurement of X-ray transitions to the ground state, providing direct access to the low-energy $K^-d$ interaction. In parallel, SIDDHARTA-2 has carried out a broad experimental programme with gaseous targets, including kaonic helium-4 and kaonic neon, as well as measurements of high-$n$ transitions in intermediate-mass kaonic atoms formed in the materials of the experimental setup. In this paper, the main SIDDHARTA-2 measurements and their physics motivations are reviewed, together with the experimental strategy developed for kaonic deuterium and the future perspectives of the collaboration within the EXKALIBUR programme.

\end{abstract}

\noindent\textbf{keywords:} kaonic atoms, X-ray spectroscopy, strong interaction, bound-state QED

\section{Introduction}
\label{sec:introduction}

Kaonic atoms are exotic atomic systems in which a negatively charged kaon ($K^-$) is electromagnetically bound to a nucleus. Because of the much larger mass of the kaon compared with that of the electron, the corresponding atomic orbits are strongly reduced and, for low-lying states, the kaon wave function significantly overlaps the nuclear region. The atomic levels are therefore affected by the strong interaction in addition to the electromagnetic interaction. These effects appear as an energy shift $\varepsilon$ and a broadening $\Gamma$ with respect to the purely electromagnetic value and can be accessed through precision X-ray spectroscopy \cite{Curceanu:2019uph}.

In this region, the strong interaction between the antikaon and the nucleus modifies the atomic levels, producing an energy shift and a finite width, with the largest effects occurring for the lowest-lying states. These deviations from the purely electromagnetic values provide direct information on the low-energy kaon--nucleus interaction and are therefore relevant for the study of non-perturbative QCD and for testing effective descriptions based on chiral SU(3) dynamics.
For hydrogenic kaonic atoms, such as kaonic hydrogen and kaonic deuterium, the strong-interaction shift, $\varepsilon_{1s}$, and width, $\Gamma_{1s}$, of the ground state are directly connected to the complex $K^-p$ and $K^-n$ scattering lengths. Kaonic hydrogen has been measured with progressively improved precision by the KpX, DEAR, and SIDDHARTA experiments \cite{Iwasaki:1997wf,DEAR:2005fdl,SIDDHARTA:2011dsy} , whereas kaonic deuterium remained experimentally inaccessible for many years because of the very low X-ray yield of the transitions to the ground state \cite{SIDDHARTA:2013ftj}. Its measurement is crucial for separating the two isospin components of the $K^-N$ interaction and therefore provides an essential constraint on the low-energy antikaon--nucleon dynamics. Measurements of heavier kaonic atoms extend these studies by probing the interaction of antikaons with multinucleon systems and by extending the investigation of strong interactions with strangeness beyond the light-kaonic atom  sector.

For high-$n$ transitions, the influence of the strong interaction is very small, so their energies are governed predominantly by electromagnetic effects. Precision X-ray spectroscopy of these transitions therefore provides a sensitive tool for testing Bound-State QED (BSQED), particularly in exotic atomic systems characterized by compact wavefunctions and sizeable vacuum-polarization contributions \cite{Sgaramella2025Neon,Paul2021ExoticAtoms,Manti2026BSQED,Clozza2026KaonicFluorine}. 

The SIDDHARTA-2 experiment at the DA$\Phi$NE collider \cite{Curceanu:2019uph,Sirghi:2023wok} takes advantage of the low-momentum kaons produced at the $\phi$-factory to perform high-precision X-ray measurements of a wide range of kaonic atoms \cite{Curceanu:2026zjg}. Its primary goal is the kaonic deuterium measurement, but the experimental programme has also included measurements with helium and neon gaseous targets, spectroscopy of intermediate-mass kaonic atoms produced in surrounding materials.

\section{The SIDDHARTA-2 apparatus at DA$\Phi$NE}
\label{sec:apparatus}

DA$\Phi$NE is an electron--positron collider operating at a center-of-mass energy of approximately \SI{1.02}{GeV}, corresponding to the mass of the $\phi$-meson resonance. The $\phi$-meson decays into a $K^+K^-$ pair with a branching fraction of about 49\%, producing nearly monochromatic charged kaons with a momentum of about \SI{127}{MeV/c} \cite{Milardi:2021khj}. The low momentum of the kaons is particularly advantageous for kaonic-atom studies because they can be efficiently slowed down and stopped in low-density gases or in thin solid targets.

The SIDDHARTA-2 apparatus is installed above the DA$\Phi$NE interaction point. A kaon trigger composed of two plastic scintillators identifies back-to-back $K^+K^-$ pairs. A shaped Mylar degrader is used to optimize the stopping distribution of the negative kaons in the target region. The central part of the setup consists of a cryogenic target surrounded by 384 Silicon Drift Detectors (SDDs)\cite{Gatti:1983wdm,Miliucci:2022lvn, Sgaramella:2022rbl}, providing a total active area of about \SI{245}{cm^2}. The SDD system was developed specifically for precision X-ray spectroscopy in the accelerator environment and combines good energy resolution with the timing information required for background suppression \cite{Sirghi:2023wok}.

\begin{figure}[htbp]
        \centering
        \includegraphics[width=0.48\textwidth]{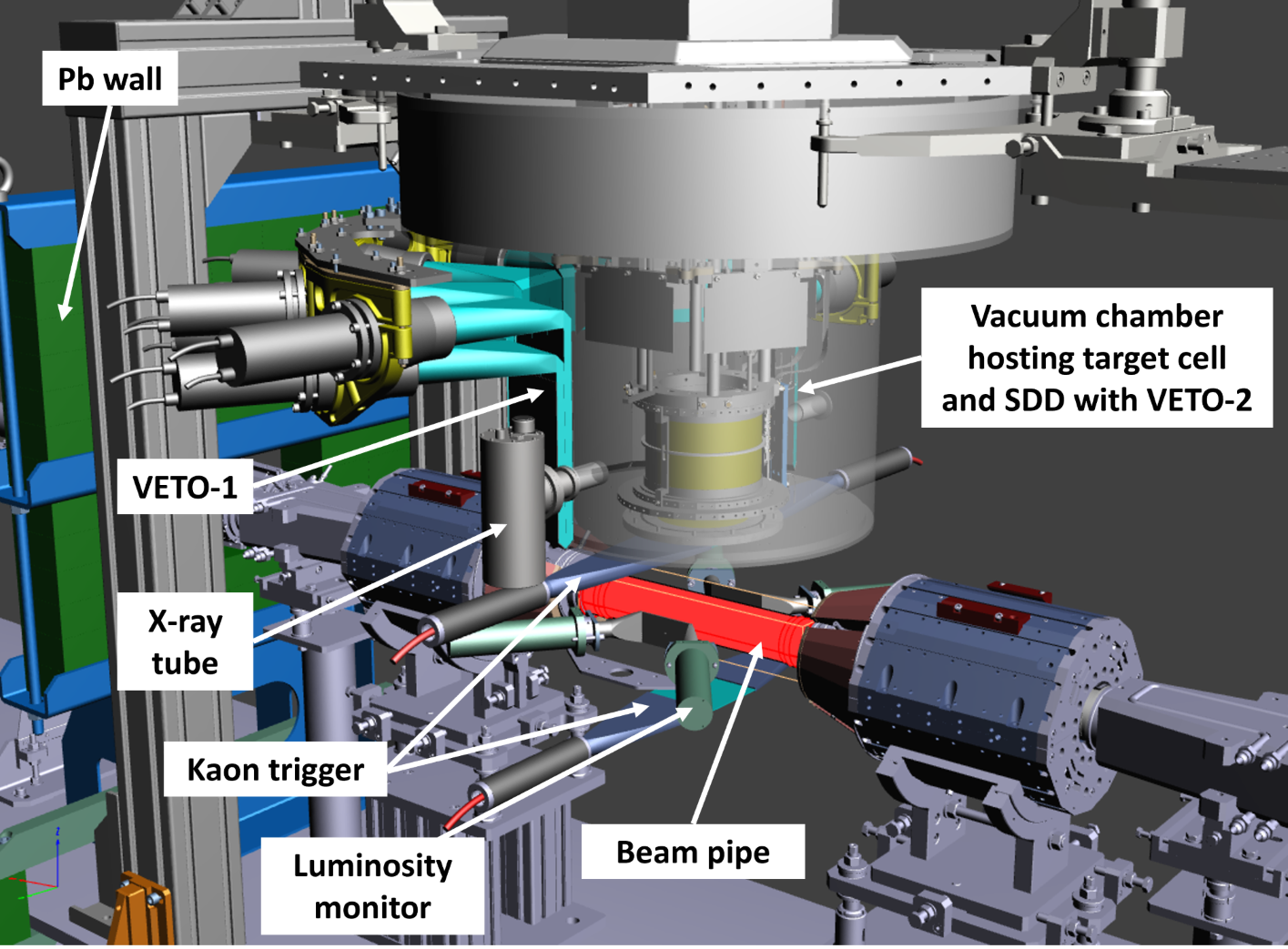}
         \includegraphics[width=0.48\textwidth]{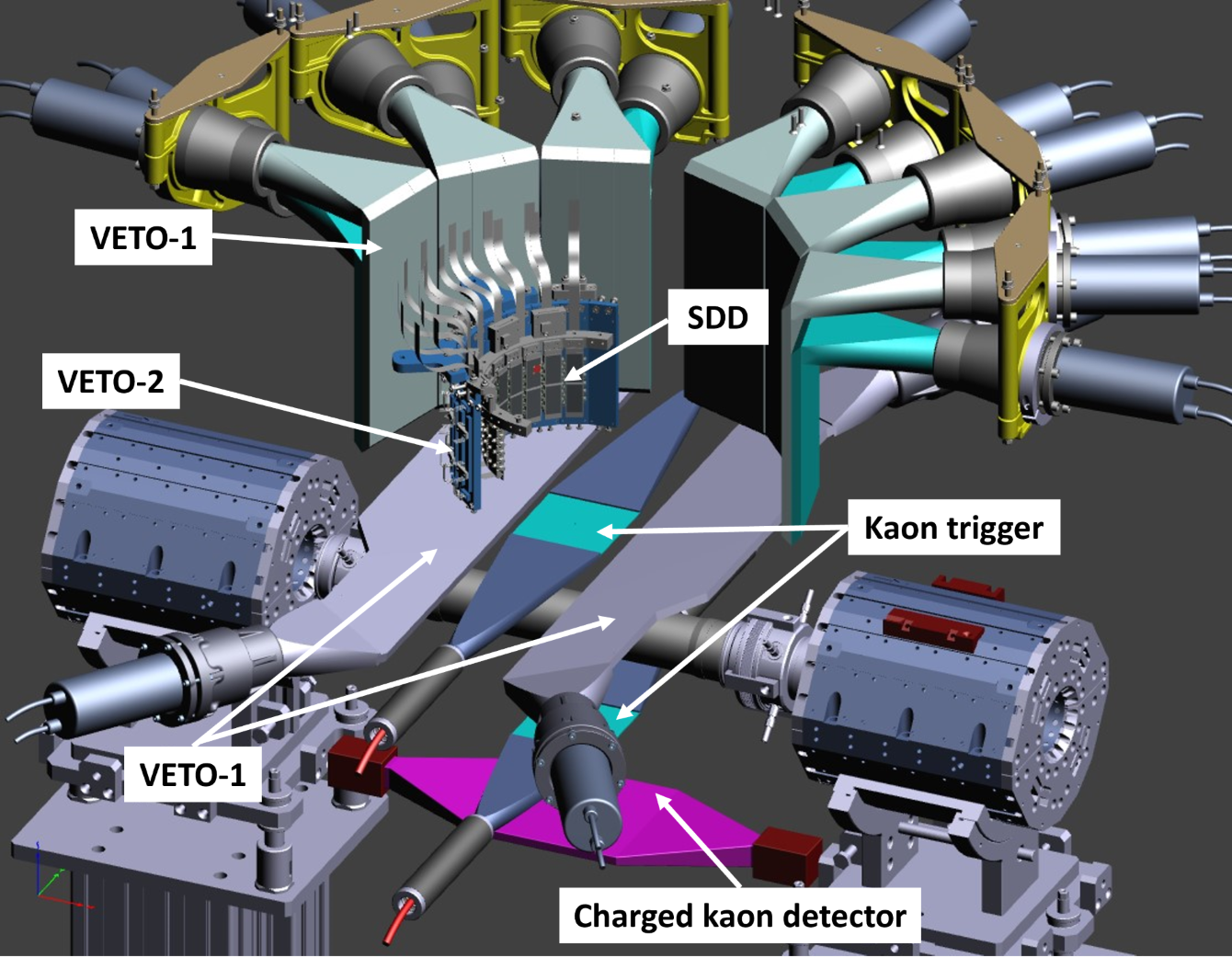}
        \caption{Left: Schematic view of the SIDDHARTA-2 experimental setup, with the main components indicated. Right: detailed view of the SDD array and the background-rejection systems, including the Veto-1 outer scintillator barrel with two additional scintillators below the vacuum chamber, the Veto-2 inner ring of plastic scintillator tiles positioned behind the SDDs, and the charged-kaon detector located below the lower scintillator of the kaon trigger. Adapted from \cite{Sirghi:2023wok}.}
        \label{fig:SIDD2_setup}
    \end{figure}

The experimental background has both electromagnetic and hadronic components. The electromagnetic component, mainly related to beam losses, is  asynchronous with kaon production and can be strongly reduced using the coincidence between the kaon trigger and the SDD timing. 

SIDDHARTA-2 includes dedicated veto detectors to suppress the hadronic background \cite{Tuchler:2023ozy,Sirghi:2023wok}. The system comprises an external scintillator barrel (Veto-1), an inner ring of scintillator tiles located close to the SDDs (Veto-2), and a kaon charge-identification detector. These detectors reduce background from charged particles produced in kaon absorption and decay, as well as X-rays originating from kaon interactions with the surrounding materials. Further details on the experimental setup can be found in \cite{Sirghi:2023wok,Skurzok:2020phi}.

\section{SIDDHARTA-2 kaonic-atom measurements}
\label{sec:measurements}

\subsection{Kaonic helium-4}
\label{sec:helium}

\begin{figure}[htbp]
    \centering
    \includegraphics[width=0.8\textwidth]{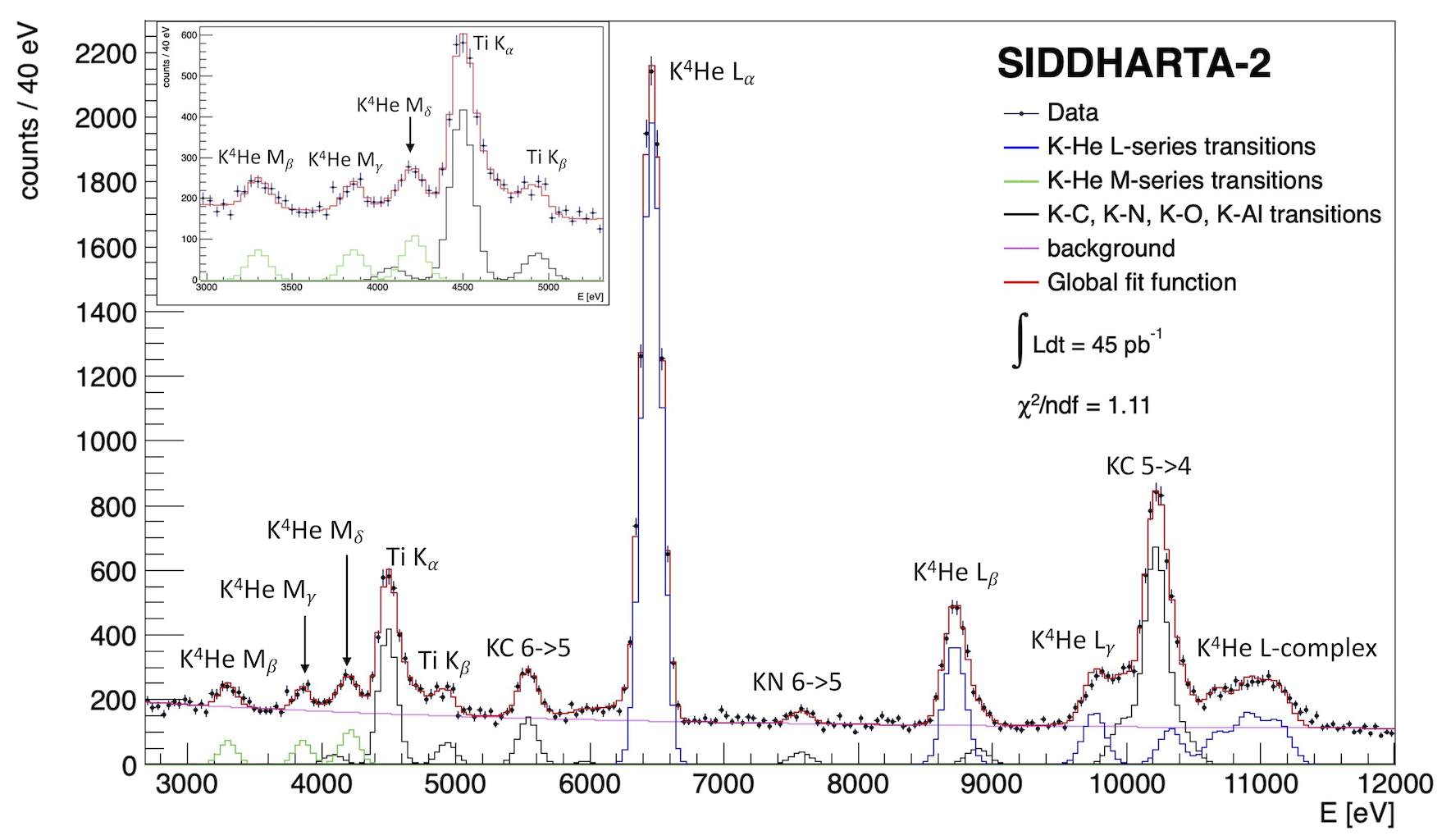}
    \caption{Kaonic helium-4 energy spectrum and fit obtained by SIDDHARTA-2 \cite{Sgaramella:2023jsc}. The kaonic helium-4 L-series and M-series transitions are highlighted.}
    \label{fig:SIDD2_KHe}
\end{figure}

Kaonic helium-4 provided an important characterization of the complete SIDDHARTA-2 setup under conditions similar to those required for the kaonic deuterium campaign, through the measurement of the kaonic helium L-lines, which are characterized by a significantly higher X-ray yield. A low-density gaseous target was used, allowing both L-series and M-series transitions to be observed \cite{Sirghi2022SIDDHARTINO,Sirghi:2023scw,Sgaramella2024Helium}. SIDDHARTA-2 measured for the first time the M-series transitions to the $3d$ level in gaseous kaonic helium-4 and determined the energy of the $L_{\alpha}$ transition with high precision \cite{Sgaramella:2023jsc}. The spectrum shown in Figure \ref{fig:SIDD2_KHe} clearly shows the kaonic helium transitions, together with background lines originating from kaon interactions with materials of the experimental setup.

The transition energies are summarized in Table~\ref{tab:helium_energies}.

\begin{table}[htbp]
\centering
\caption{The SIDDHARTA-2 kaonic helium-4 transition energies measured with a gaseous target by SIDDHARTA-2 \cite{Sgaramella:2023jsc}. The first uncertainty is statistical and the second systematic.}
\label{tab:helium_energies}
\begin{tabular}{lc}
\toprule
Transition & Energy [eV] \\
\midrule
$L_{\alpha}\;(3d\rightarrow2p)$ & $6461.4 \pm 0.8 \pm 2.0$ \\
$M_{\beta}\;(5f\rightarrow3d)$  & $3300.8 \pm 13.2 \pm 2.0$ \\
$M_{\gamma}\;(6f\rightarrow3d)$ & $3860.4 \pm 13.6 \pm 2.2$ \\
$M_{\delta}\;(7f\rightarrow3d)$ & $4214.1 \pm 19.6 \pm 2.2$ \\
\bottomrule
\end{tabular}
\end{table}

The strong-interaction shift and width of the $2p$ level were determined to be
\begin{equation}
\varepsilon_{2p}=-1.9\pm0.8_{\rm stat}\pm2.0_{\rm syst}\;\mathrm{eV},
\end{equation}
and
\begin{equation}
\Gamma_{2p}=0.01\pm1.60_{\rm stat}\pm0.36_{\rm syst}\;\mathrm{eV},
\end{equation}
showing no significant strong-interaction effect at the achieved precision.

The helium measurement also provided information on the atomic cascade through the X-ray yields. The published SIDDHARTA-2 yields and yield ratios at the same target density are collected in Table~\ref{tab:helium_yields}.

\begin{table}[htbp]
\centering
\caption{Kaonic helium-4 X-ray yields per stopped kaon of the M-series and L-series transitions measured at a density  of $1.37\pm0.07$~g/l \cite{Sgaramella:2023jsc}.}
\label{tab:helium_yields}
\begin{tabular}{lc}
\toprule
Transition & Yield \\
\midrule
$L_{\alpha}$ yield & $0.119\pm0.002{}^{+0.006}_{-0.009}$ \\
$M_{\beta}$ yield  & $0.026\pm0.003{}^{+0.010}_{-0.001}$ \\
$L_{\beta}/L_{\alpha}$ & $0.172\pm0.008$ \\
$L_{\gamma}/L_{\alpha}$ & $0.012\pm0.001$ \\
$M_{\beta}/L_{\alpha}$ & $0.218\pm0.029$ \\
$M_{\gamma}/M_{\beta}$ & $0.48\pm0.11$ \\
$M_{\delta}/M_{\beta}$ & $0.43\pm0.12$ \\
\bottomrule
\end{tabular}
\end{table}

These measurements provide experimental input for kaonic-atom cascade models and demonstrate the capability of SIDDHARTA-2 to perform precision spectroscopy with low-density gaseous targets.

\subsection{Kaonic neon}
\label{sec:neon}

\begin{figure}[htbp]
    \centering
    \includegraphics[width=0.7\textwidth]{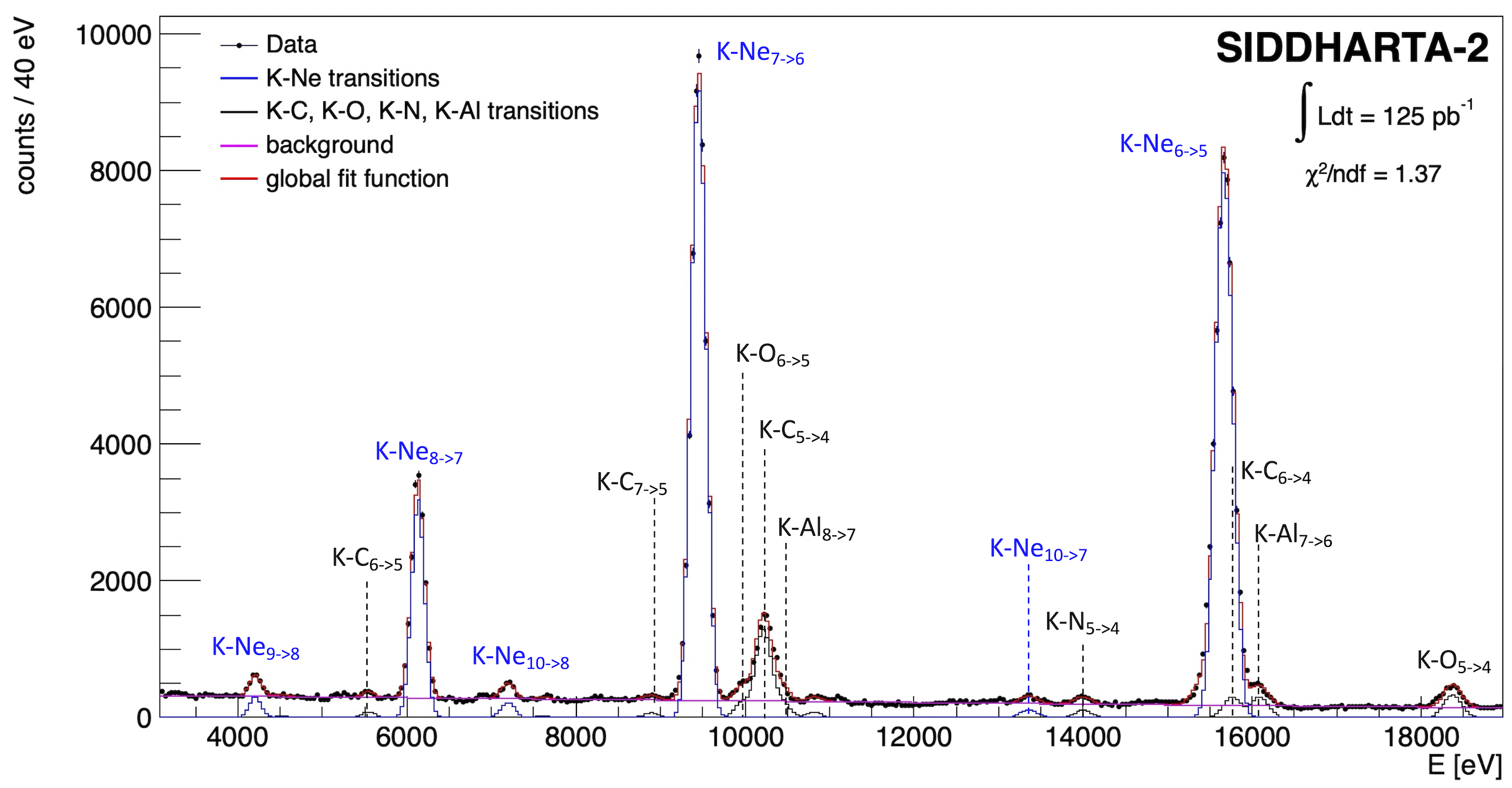}
    \caption{Kaonic neon energy spectrum and corresponding fit obtained by the SIDDHARTA-2 collaboration. \cite{Sgaramella2025Neon}. }
    \label{fig:KNe}
\end{figure}

SIDDHARTA-2 also performed the first high-precision kaonic-neon X-ray spectroscopy with a gaseous target \cite{Sgaramella2025Neon}. The target density was $3.60\pm0.18$~g/l and the data set corresponded to an integrated luminosity of \SI{125}{pb^{-1}}. Six high-$n$ transitions were clearly identified, and three of them were measured with sub-eV statistical uncertainty (see Figure \ref{fig:KNe}).

Table~\ref{tab:neon} summarizes the transition energies and yields per stopped kaon.

\begin{table}[htbp]
\centering
\caption{SIDDHARTA-2 kaonic-neon transition energies and yields per stopped kaon \cite{Sgaramella2025Neon}. For both energy and yield, the first uncertainty is statistical and the second systematic.}
\label{tab:neon}
\small
\begin{tabular}{lcc}
\toprule
Transition & Energy [eV] & Yield \\
\midrule
$10\rightarrow8$ & $7191.21\pm4.91\pm2.00$ & $0.010\pm0.001\pm0.001$ \\
$10\rightarrow7$ & $13352.20\pm10.07\pm3.00$ & $0.004\pm0.002\pm0.001$ \\
$9\rightarrow8$  & $4206.35\pm3.75\pm2.20$ & $0.137\pm0.012\pm0.010$ \\
$8\rightarrow7$  & $6130.86\pm0.71\pm1.50$ & $0.228\pm0.004\pm0.011$ \\
$7\rightarrow6$  & $9450.08\pm0.41\pm1.50$ & $0.277\pm0.002\pm0.014$ \\
$6\rightarrow5$  & $15673.30\pm0.52\pm9.00$ & $0.308\pm0.003\pm0.015$ \\
\bottomrule
\end{tabular}
\end{table}

The achieved precision demonstrates the potential of kaonic neon for future BSQED studies. The use of a low-$Z$, low-density gaseous target also reduces electron-screening and electron-refilling effects, making kaonic neon a promising system for a future high-precision determination of the charged-kaon mass.

\subsection{Intermediate-mass kaonic atoms}
\label{sec:intermediate}

Using data collected during the DA$\Phi$NE commissioning phase, SIDDHARTA-2 measured several high-$n$ transitions of intermediate-mass kaonic atoms in the 5-20 keV energy range \cite{Sgaramella:2023orc}. The observed lines originate mainly from kaons stopping in solid materials of the setup, such as the aluminum frame and the Kapton target window. Transitions from kaonic carbon, nitrogen, oxygen, and aluminum were measured with a precision of a few eV. The transition energies are summarized in Table~\ref{tab:intermediate}.

\begin{table}[htbp]
\centering
\caption{High-$n$ transition energies of kaonic carbon, nitrogen, oxygen, and aluminum measured by SIDDHARTA-2 \cite{Sgaramella:2023orc}. The first uncertainty is statistical and the second systematic.}
\label{tab:intermediate}
\small
\begin{tabular}{lll}
\toprule
Atom & Transition & Energy [eV] \\
\midrule
K$^-$C  & $6\rightarrow5$ & $5541.7\pm3.1\pm2.0$ \\
        & $7\rightarrow5$ & $8890.0\pm13.0\pm2.0$ \\
        & $5\rightarrow4$ & $10216.6\pm1.8\pm3.0$ \\
        & $6\rightarrow4$ & $15760.3\pm4.7\pm12.0$ \\
        & $7\rightarrow4$ & $19099.0\pm32.0\pm9.0$ \\
\addlinespace
K$^-$O  & $7\rightarrow6$ & $6014.8\pm8.4\pm2.0$ \\
        & $6\rightarrow5$ & $9968.1\pm6.9\pm2.0$ \\
        & $7\rightarrow5$ & $15970.0\pm23.0\pm9.0$ \\
        & $5\rightarrow4$ & $18353.1\pm11.0\pm9.0$ \\
\addlinespace
K$^-$N  & $6\rightarrow5$ & $7577.0\pm17.0\pm2.0$ \\
        & $5\rightarrow4$ & $14010.6\pm8.2\pm9.0$ \\
\addlinespace
K$^-$Al & $8\rightarrow7$ & $10441.0\pm8.5\pm3.0$ \\
        & $7\rightarrow6$ & $16083.4\pm3.8\pm12.0$ \\
\bottomrule
\end{tabular}
\end{table}

These measurements extend the experimental kaonic-atom database and demonstrate that the SIDDHARTA-2 apparatus can perform precision spectroscopy over a broad range of atomic masses. They also provide useful input for kaon--nucleus interaction models and for future BSQED studies in exotic atoms.




\subsection{The kaonic deuterium programme}
\label{sec:deuterium}

Kaonic deuterium is the principal physics goal of SIDDHARTA-2. Its measurement is exceptionally difficult because the X-ray yield for transitions to the $1s$ level is very small and the strong-interaction broadening is expected to be large. As a consequence, the signal is weak and distributed over broad spectral structures, requiring an optimized target density, high-efficiency X-ray detection, and strong suppression of the accelerator background.

The kaonic deuterium campaign was carried out with a cryogenic gaseous target during the 2023--2024 data-taking period. The apparatus configuration was optimized specifically for this measurement, including the kaon trigger, the SDD timing, and the veto systems. The purpose of the measurement is to determine the strong-interaction shift and width of the $1s$ level, which provide access to the $K^-d$ interaction at threshold and, together with kaonic-hydrogen information, to the isospin-dependent antikaon--nucleon scattering lengths.

 A preliminary spectrum was obtained using a subset of the total data sample, corresponding to about $164~\mathrm{pb}^{-1}$ out of the approximately $1~\mathrm{fb}^{-1}$ collected for the kaonic deuterium measurement. This preliminary analysis was used to optimize the analysis strategy and to assess the performance of the experimental setup. Assuming a kaonic deuterium $1s$ energy shift in the range from $-1000$ to $-600$ eV and a width of about 1 keV, the expected region of interest for the $K_{\alpha}$ transition is approximately 6.6-7.4 keV. A signal structure is visible within the selected region of interest, while a more detailed analysis of the complete dataset is available \cite{Sgaramella_kd2026}.

\begin{figure}[htbp]
        \centering
        \includegraphics[width=0.6\textwidth]{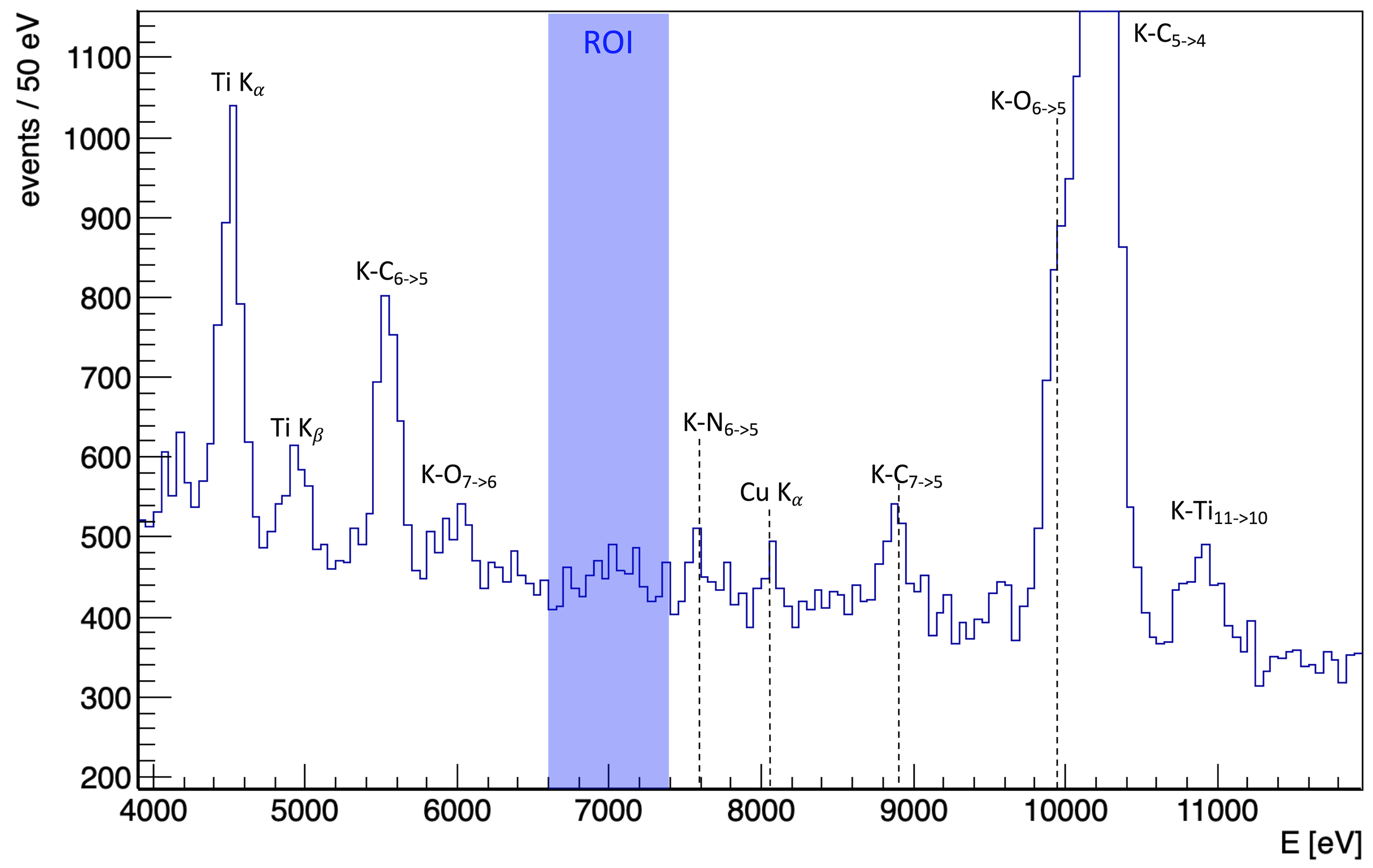}
        \caption{Preliminary SIDDHARTA-2 energy spectrum obtained from a subset ($164~\mathrm{pb}^{-1}$)  of the total data sample collected for the kaonic deuterium measurement, after background subtraction. The region of interest (ROI) for the kaonic deuterium K$\alpha$ line is highlighted. }
        \label{fig:Kd_spectrum}
    \end{figure}

\section{Future perspectives: EXKALIBUR}
\label{sec:future}

The EXKALIBUR programme (EXtensive Kaonic Atoms research: from LIthium and Beryllium to URanium) builds on the developments achieved with SIDDHARTA-2 and aims to extend high-precision kaonic-atom spectroscopy over a broad range of the periodic table \cite{Curceanu:2023yuy,Sgaramella2026EXKALIBUR}. Its first phase will exploit the SIDDHARTA-2 setup for precision measurements of kaonic neon, light kaonic atoms such as lithium, beryllium, and boron, and intermediate-mass systems including oxygen, aluminum, and sulfur. A new generation of \SI{1}{mm}-thick SDDs is being developed to improve the detection efficiency at higher X-ray energies, together with an upgraded calibration strategy.
A major objective of this programme is a new determination of the charged-kaon mass using purely electromagnetic kaonic-atom transitions. The value presently reported by the Particle Data Group is obtained from a weighted average of the two most precise measurements, which are in significant disagreement. Both measurements were performed with solid targets and are affected by systematic uncertainties related to electron screening and electron refilling. These effects are strongly reduced in low-Z gaseous targets. For this reason, a dedicated high-precision measurement with gaseous kaonic neon is planned, taking advantage of the accuracy already demonstrated by SIDDHARTA-2 for high-$n$ transitions. Such a measurement could provide an independent and more precise determination of the charged-kaon mass.
More generally, the EXKALIBUR programme will address several complementary physics topics, including kaon--multinucleon interactions, Bound-State QED, and systematic studies of kaonic atoms across the periodic table.  Future measurements are being considered at both DA$\Phi$NE and J-PARC, taking advantage of the complementary capabilities of the two facilities to extend precision kaonic atom spectroscopy across the periodic table.

\section{Conclusions}
\label{sec:conclusions}

Kaonic atoms provide a unique tool for investigating low-energy strong interactions in systems with strangeness and, through high-$n$ transitions, for testing Bound-State QED. Within this framework, SIDDHARTA-2 has established a broad programme of high-precision X-ray spectroscopy at the DA$\Phi$NE collider, exploiting low-momentum kaons produced in $\phi$-meson decays.
The measurements performed with gaseous helium-4 and neon targets have demonstrated the excellent performance of the experimental setup and have provided new information on kaonic-atom cascades, strong-interaction effects, and high-$n$ transitions relevant for BSQED studies. Additional measurements of intermediate-mass kaonic atoms, including carbon, nitrogen, oxygen, and aluminum, have further extended the experimental programme and shown the versatility of the SIDDHARTA-2 approach.
The kaonic deuterium campaign remains the central and most challenging measurement of SIDDHARTA-2, with the full data set currently under analysis. The experience and developments achieved with SIDDHARTA-2 now provide the basis for the EXKALIBUR programme, which will extend precision kaonic-atom spectroscopy to a wider range of nuclear targets, with the aim of systematically investigating kaon--nucleus interactions, BSQED effects, and the charged-kaon mass.

\section*{Acknowledgments}

We thank C. Capoccia from LNF-INFN and H. Schneider, L. Stohwasser, and D. Pristauz-Telsnigg from Stefan Meyer-Institut for their fundamental contribution in designing and building the SIDDHARTA-2 setup. We gratefully acknowledge the Polish high-performance computing infrastructure PLGrid (HPC
Center: ACK Cyfronet AGH) for providing computer
facilities and support within the computational grant
no. PLG/2025/018524. We also thank INFN-LNF and
the DA$\Phi$NE staff for the excellent working conditions
and their ongoing support. Special thanks to Catia
Milardi for her continued support and contribution
during the data taking. Part of this work was supported by the INFN (KAONNIS project); the Austrian
Science Fund (FWF): [P24756-N20 and P33037-N];
the Croatian Science Foundation under the project
IP-2022-10-3878; the EU STRONG-2020 project (Grant
Agreement No. 824093); the EU Horizon 2020 project
under the MSCA (Grant Agreement 754496); the
Japan Society for the Promotion of Science JSPS
KAKENHI Grant No. JP18H05402, JP22H04917;
the Polish Ministry of Science and Higher Education grant No. 7150/E-338/M/2018 and the Polish
National Agency for Academic Exchange (grant no
PPN/BIT/2021/1/00037). This article/publication is
based upon work from COST Action CA24131 (ENRICH),
supported by COST (European Cooperation in Science
and Technology, http://www.cost.eu/).

\bibliography{iopart-num}

\end{document}